\documentstyle[aps,prb,multicol,epsf,epsfig]{revtex} 
 
\newcommand{\be}{\begin{equation}}   
  \newcommand{\ee}{\end{equation}}   
\newcommand{\bea}{\begin{eqnarray}}   
  \newcommand{\eea}{\end{eqnarray}}   
\def\sig{{\boldmath$ \sigma$}}   
   
\begin{document}   
  
\title{  Geometry Selects Highly Designable Structures}   
\author{ V. Shahrezaei$^{1,2}$ and  M.R. Ejtehadi$^{1,3}$}   
\address{$^1$ {\it Institute for studies in Theoretical Physics and   
    Mathematics,   
    Tehran  P.O. Box 19395-5531, Iran.}\\   
  $^2$ {\it Department of Physics, Sharif University of Technology,   
    Tehran P.O. Box: 11365-9161, Iran.}\\   
  $^3$ {\it Max-Planck Institut f\"ur    
    Polymerforschung, Ackermannweg 10, D-55128 Mainz, Germany  }\\   
  }    
  
\maketitle    
 
\begin{abstract}    
  By enumerating all sequences of length $20$, we study the designability of   
  structures in a two-dimensional Hydrophobic-Polar (HP) lattice model in a   
  wide range of inter-monomer interaction parameters. We find that 
  although the    
  histogram of designability depends on interaction parameters,    
  the set of highly designable structures is invariant. So in the HP
  lattice model the High
  Designability should be a purely  
  geometrical feature.   
  Our results suggest two geometrical properties for highly  
  designable structures, they have maximum number of contacts and unique 
  neighborhood    
  vector representation. Also we show that contribution of perfectly
  stable sequences in designability of structures plays a major role
  to make them highly designable.    
\end{abstract}   
  
\begin{multicols}{2}   
    
  Proteins are bio-macromolecules, which consist of linear sequences of   
  monomers, the 20 naturally occurring amino acids.   
  Each protein folds to a unique spatial structure as its  
  native state, which   
  is its global minimum of the free energy \cite{Anfinsen}. This
  structure specifies   
  the functionality of the protein sequence in the nature.   
 
  It has been noted that certain structures are more commonly     
  observed among proteins than others \cite{Chothia,Orengo}.   
  There are efforts to explain this phenomenon by   
  considering protein structure designability, defined as the number of   
  sequences that would successfully fold in one structure \cite{Tang}.
  By definition,  
  highly designable structures (HDSs) have more chance to be found as native,   
  and also they have good stability against sequence mutations. 
  A recent numerical analysis on experimental data revealed that the
  distribution of observed protein families over different folds can be
  modeled with a highly stretched exponential \cite{Govindara}. This
  observation is quite 
  consistent with designability explanation. Highly designable structures might 
  also represent attractive targets for protein design \cite{Hellinga}.

  In study of designability of native structures like many other
  features in the field of protein folding, the complexity of the
  problem forces to use more simplified models. Coarse grained view
  point to proteins introduce effective inter-monomer interactions as
  a relevant factor in designability of structures. Analyzing the
  interaction between the 20 amino acids suggests that amino asides
  can be separated to Hydrophobic (H) and Polar (P) groups \cite{Li0}. This
  introduces a very simple but highly popular two monomer types model,
  HP model. Although the chains in  this simple model are too far from
  realistic protein and surly could not explain many features of real
  proteins, but this model had good success in clarifying the concept
  of designability  
  \cite{Li}. The
  simplicity of this model allows one to study the ground state
  properties of model proteins by enumerating chains and configurations for
  short length chains. 

  Enumerations on two and three dimensional lattice models have shown
  the existence of a few highly designable structures among many lowly  
  designable ones \cite{Li,Chan,Bornberg}.     
  Some studies on dynamical properties of those model chains, which fold  
  in the highly    
  designable lattice structures, show that they are more protein-like. Such   
  sequences are thermodynamically more stable \cite{Li} and they fold to   
  native state faster than   
  random sequences \cite{Goldstein,Melin}.    

  There have been many efforts to find the factors determining the
  high designability of a structure using simple models. One element
  is the set of 
  inter-monomer interactions   
  \cite{Pande,Self1,Kussel}. Some evidences suggest      
  that the set of highly designable structures depends on the number   
  of monomer types in the model \cite{Buchler}. However in 
  lattice models with short chains, it is too difficult to talk about
  helixes and sheets, it is claimed that  
  HDSs possess secondary     
  structure in two and three dimensional HP models \cite{Li}. By the
  use of a clever 
  algebraic approach    
  in the framework of a simple solvation model it is shown that HDSs should    
  be rare and atypical in structure space \cite{Li2}.  A recent argument    
  shows that this result remains valid for more general models  
  \cite{Buchler2}.     
    
  In this paper we study the designability of lattice structures in a  
  wide range of interaction parameters between H and P monomers by using of a
  recently developed method  \cite{Self2}. Results confirm that   
  designability of structures depends on inter-monomer interactions but   
  interestingly we find that the set of HDSs is invariant. Therefore   
  in this simple model the geometry should have the essential role in
  the selection of some    
  structures as most HDSs. Also the correlation with some other geometrical   
  properties will be shown.

  We use of a   
  two dimensional Hydrophobic-Polar (HP) lattice Model \cite{Chan2} for   
  sequences with length 20. It is obvious that two dimensional structures
  have significant differences with real three dimensional ones, but to
  get considerable results in three dimension it needs to go to the
  too longer chains which is not computationally accessible. Short chains in 
  three dimension do not possess natural ratio of core sites. The method of this 
  paper is based on contact matrix. Thus it 
  can be easily generalized to any pair contact model. 
  
  In Pair Contact Models the energy of a given sequence    
  $\sigma$ in a given structure can be written as    
  \be    
  \label{1.0}    
  E=\sum_{i,j} c_{ij} m_{\sigma_i \sigma_j}.     
  \ee    
  Where $c_{ij}$ and $m_{\sigma_i\sigma_j}$   
  are respectively the elements of the contact matrix ($C$) and the   
  interaction matrix ($M$). $c_{ij}$ is 1 if the monomers $i$ and $j$ are   
  non-sequential neighbor and is 0 otherwise. $\sigma_i$ is  
  $i$th component of sequence vector $\sigma$. In our HP model, it is 
  equal to $0$   
  ($-1$) if $i$th   
  monomer in sequence is a polar (hydrophobic) residue.  The 
  $m_{\sigma_i\sigma_j}$ is the    
  interaction energy between monomer type $\sigma_i$ and $\sigma_j$. 
    
  In two-dimensional square lattice there are $41,889,578$ distinct 
  structures    
  (non-related by rotation or reflection symmetries) for a sequence with   
  length 20.  There is a contact matrix corresponding to each structure. It   
  is possible that some structures have the same contact matrix. Such   
  contact matrices which point to more than one structure are called   
  degenerate contact matrices. The number of non-degenerate distinct  
  contact matrices are   
  about 1 million which is much less than the number of all possible  
  structures.     
  The maximum number of possible contacts for the   
  sequences with length 20 is 12, and the number of maximally compact   
  structures i.e. with maximum contacts, are 503.   
    
  In HP model the monomers are divided to two Hydrophobic (H) and   
  Polar (P) groups. The interaction matrix $M$ is thus a $2\times2$ matrix   
  and its elements are $E_{HH}$ and $E_{HP}$ and $E_{PP}$. By choosing   
  an arbitrary energy scale, we can parameterize these three elements of   
  the interaction matrix in terms if two positive parameters,   
  $\gamma$ and $E_c$ \cite{Self2}.     
  \bea   
  \label{2.0}    
  E_{HH} &=& -2 -\gamma -E_c, \nonumber\\    
  E_{HP} &=& -1 - E_c, \nonumber \\    
  E_{PP} &=& - E_c,    
  \eea    
  Substituting these elements of interaction matrix in equation  
  (\ref{1.0}) gives following simple expression for configuration energy.    
  \be    
  \label{3.0}    
  E= -m -E_c \cdot b - \gamma \cdot a    
  \ee    
  where $m$, $b$ and $a$ are   
  three positive   
  integers, related to $\sigma$ and $C$ as follows:    
  \bea   
  \label{4.0}   
  m&=& - {\mbox {\sig}}^{t} \cdot C \cdot{\bf 1}, \nonumber \\   
  a&=&  {1\over2} {\mbox{\sig}}^{t} \cdot C   
  \cdot{\mbox{\sig}}, \nonumber \\   
  b&=& {1\over2} {\bf 1}^{t} \cdot C \cdot{\bf 1}.   
  \eea

  Using the set of possible ground states of sequences,   
  enables us to find the ground state of any sequence for any given   
  values of energy parameters \cite{Self2,Self3}. In this way the  
  designability of all    
  structures in a wide range of energy parameters $E_c$ and $\gamma$ is   
  obtained by enumerating the all $2^{20}$ sequences. Fig 1 shows the   
  average designability of structures in a 10 by 10 square region with the   
  mesh $0.1$ in the space of $E_c$ and $\gamma$. In this average    
  the structures with zero designability are excluded. 19,132   
  structures (about $0.05\%$ of all structures) have non-zero  
  designability at least for     
  some given values of   
  $E_c$ and $\gamma$.  As one can see in this figure, the average   
  designability shows two different regimes in space of energy parameters.   
  In a wide area of energy parameters it has a value in order of $10$, but   
  for large $E_c$ and small $\gamma$ it rapidly jumps to several hundreds.   
  This can be explained by the fact that for $E_c \gg \gamma$ the   
  contribution of $b$ in the configuration energy (eq. \ref{3.0}) is more   
  essential. Therefore, all native structures are between the most   
  compact ones. We can call this area of interaction parameters, {\it  
    compact regime}.    
  In this regime reduction in the number of native structures, increases   
  the average designability. Alternatively when $\gamma$ is greater than   
  $E_c$ or comparable with it, some non-compact configuration  
  can compete with compact ones. We call this area {\it swollen  
    regime}.    
    
  Fig. 2 shows the histogram of designability for two pairs of   
  values of $E_c$ and $\gamma$.  Fig 2.a is for the  
  pair $E_c=3$, $\gamma=8$ which is a point in the swollen regime and 
  fig 2.b is for the pair $E_c=9$, $\gamma=0.5$   
  which is a point in the compact regime . In  
  the swollen  regime   
  where native structures are not restricted to only highly compact  
  structures, there are many lowly designable structures and a few  
  HDSs. Alternatively  in the compact Regime   
  there are many intermediately designable structures and again a few 
  HDSs. Similar results are reported when one compares histograms of 
  designability for a fixed given set of energy parameters using the search 
  space of all structures and compact structures \cite{Tatsumi}.   
  Therefore, the interaction parameters just choose the search space  
  and inside each regime (compact or  
  swollen), the statistics of designability does not change qualitatively.       
   
  The relevant question is, if the set of HDSs depend on the interaction  
  parameters.   
  To study this, we sort the structures by their   
  designabilities at any set of energy   
  parameters. In this manner, the place of any structure in competition for   
  designability is recognized with its rank. The rank one is the most   
  HDS for those given energy parameters. Averaging the rank of one   
  structure in all space of energy parameters gives a good perspective about   
  its global attitude toward high designability.   
   
  Figure 3 shows the histogram of this average rank for all structures in   
  the studied square region of energy parameters. The interesting  
  point in this    
  diagram is that there are a few structures with very small average rank.   
  For example there are $8$ structure   
  with average rank less than 10 and lowest average rank value is equal to   
  $1.48$. It must be noted that since the rank of a structure is a positive 
  quantity, the smallness of its average shows that the structure has
  small rank    
  in all space of energy parameters. Thus structures with very low rank  
  always are inside the hit-list of HDSs.    
  We compared the behavior of average rank of   
  structures with two asymptotic limits. If the rank of structures were   
  invariant against the change of energy parameters, the histogram would   
  show a completely flat behavior (solid line in figure 3). In the other   
  side, if the structures rank were changed uncorrelatively by the  
  changing the inter-monomer interactions, it would result a very sharp  
  Gaussian    
  distribution as a consequence of central limit theorem (dashed line in   
  figure 3). As one can see in figure 3 it seems that the average rank   
  behaves more similar to the case of quenched rank, than the random case.   
  Especially, for the low ranks the histogram lies on the solid line very   
  well. This shows that the ranks of HDSs are more rigid than the ranks of   
  lowly designable structures. So, the set of HDSs does not depend on the   
  interaction parameters, although the designability of structures does.   
  The fact that the role of the interaction parameters is not important for   
  choosing a structure as HDS, demonstrates the importance of geometry.   
  This leads to the question which purely geometrical properties or
  symmetries
  select some structures as HDS. Furthermore it justifies the
  restriction of the search for HDS to highly compact structures.   
    
  Recently we have shown in HP lattice model that there are some 
  sequences which    
  have only one non-degenerate possible ground state in all space of energy   
  parameters \cite{Self2}. Because the native structures of these sequences   
  have perfect stability against the changing of inter-monomer interaction,   
  we call them perfectly stable sequences (PSSs). PSSs give constant   
  contribution to the designability of structures. For any given interaction   
  parameters, each set of sequences which have a common ground state
  structure     
  contains an invariant subset, constituted by PSSs. The   
  designability of any structure has a {\it constant part} equal to the   
  number of the members of its invariant subset of sequences.  In our model,   
  about $7\%$ of the sequences of length $20$ are PSS.  These PSSs select   
  489 structures as their absolute native state out of the 503 most compact    
  structures i.e. only 489 structures have a non-zero constant part of   
  designability.(It is not possible that PSSs select non-compact   
  structures as ground state because with a large enough $E_c$    
  the compact structures will gain lower energies.)   
     
  There is a strong correlation between the   
  designability and its constant part. In figure 4 we have plotted the
  constant part of designability against the designability, for two
  sets of interaction parameters used in figure 2. This figure shows
  that designability is nearly    
  proportional to its constant part, but the slope is a function of   
  interaction parameters.   
   
  Indeed, in the swollen regime the slope is of
  order 
  unity (fig 4.a). This means that the invariant subsets of sequences  
  dominate the designability of structure.   
  But in the compact regime, where the  
  constant part of    
  designability has a small contribution to designability, these quantities   
  remain nearly proportional yet, and the most HDSs have the bigger constant   
  parts (fig 4.b). This correlation is considerable because by
  definition the source of    
  designability is different from its constant part. The former shows    
  stability against monomer mutations in the sequence, and the later shows   
  ability of structure to be an interaction independent native state.   
  The strong correlation between these quantities suggests   
  that they may have the same geometrical origin.    
  It has been reported that designability has a good correlation with   
  Energy gap \cite{Li}. So it can be claimed that the existence of  
  a large constant part in     
  designability of a structure is a sign of large average energy
  gap. In fact, when a PSS exist, changing interaction parameters does 
  not change the native    
  structure of sequence. So the exited states of the sequence should  
  be far enough  and   
  separated by a large energy gap.     
   
  As mentioned above the compactness is a necessary condition of HDSs
  because all interaction parameters are negative.    
  We have found an additional geometrical symmetry for HDSs which is  
  related to the
  solvation nature of proteins. In a HP solvation   
  model for proteins a H monomer decrease the energy proportional to  
  the number of     
  non-sequential contacts in the structure and the  
  position of P monomers does
  not change the energy. In a simpler version of the solvation model all  
  non-core monomers behave  the same (it takes both corner and edge monomers  
  as surface.)  
  \cite{Li2,Buchler2,Shih}.  
  In contrast the solvation model is a     
  pair contact model. Our pair contact model in the special case
  $\gamma=0$ becomes    
  a solvation model. (We may call it alternatively {\it additive} 
  \cite{Self1}.) 
  This special case has some theoretical advantages because the energy has a 
  simpler form \cite{Self1,Kussel}.     
  In this case to calculate the configuration energy (eq. \ref{3.0})  
  $a$ is an irrelevant     
  parameter, and one can re-write $m$ and $b$ (equation   
  \ref{4.0}) as follows.     
  \bea    
  \label{5.0}    
  m&=& - {\mbox {\sig}}^{t} \cdot V, \nonumber \\    
  b&=& {1\over2} {\bf 1}^{t} \cdot V,     
  \eea    
  where $V$ is {\it neighborhood vector} (NV). Its $i$th component
  shows the number of   
  non sequential neighbors of the $i$th monomer and is related to the  
  contact matrix.     
  \be    
  V_i = C \cdot{\bf 1} = \sum_j c_{ij}.     
  \ee    
  In this case the information needed to  
  calculate the configuration energy can be coded in a vector instead   
  of a matrix. Obviously this is a source of an additional degeneracy in   
  energy spectrum. This degeneracy is equal to the number of spatial
  structures which have the same NV and we label it by $N_d$.    
  In our enumeration for native configurations of length $20$ the 
  biggest $N_d$ for  
  NVs is $4$.  
 
  We found that all HDSs are between those   
  structures with unique NVs ($N_d=1$). Thus this additional   
  geometric symmetry is another common property of HDSs.

  \begin{center}  
    \parbox{7cm}{\small Table 1. The average designability of
      structures } \\[3mm]  
    
    \begin{tabular}[h]{|c||c|c|c|c|}  
      \hline  
      &\multicolumn{4}{c|}{Number of contacts} \\ \cline{2-5} 
      $N_d$ & 9 &10 & 11& 12 \\  
      \hline \hline 
      1 & 0.85 & 2.07 & 14.17 & 692.78 \\ \hline  
      2 &      & 1.09 &  4.97 & 155.71  \\ \hline 
      3 &      & 1.87 &  2.65 & 101.17  \\ \hline  
      4 &      &      & 2.69  & 40.04  \\ \hline  
    \end{tabular}

  \end{center}

  Table 1 shows the average designability of structures with specific   
  compactness and degeneracy of NVs. It can be inferred   
  that the average designability of structures increase with the growth of   
  compactness and with the decreasing of $N_d$. Thus those structures which   
  have the highest compactness and unique NVs   
  are good candidates for being highly designable.   
  These are only necessary conditions for high designability and    
  as some lowly   
  designable structures also fulfill these criteria.    
  In fact the structures should be also atypical to be HDS \cite{Li2}.   
  Recently it was shown that these atypical structures possess
  $\alpha$ helices    
  which are another character of real proteins \cite{Shih}.  
  Also the uniqueness of NV representation  
  of structures can give an explanation for the   
  ratio of surface to core monomers in real proteins \cite{Shih}.  
  
  In summary we have studied the designability of structures in a   
  two-dimensional HP pair contact lattice model in a wide range of   
  inter-monomer interaction parameters by considering HP 
  constraints. We find the designability
  of all structures by  
  enumerating all sequences of length 20.    
  Our results confirm that    
  changing the inter-monomer interactions affects the structure 
  designability    
  and also chooses the search space of native state but the set of HDSs 
  is invariant.    
  Therefore geometry should have the essential role in the selection of some   
  structures as most HDSs.
  
  In some regions of inter-monomer interaction parameter
  space, the constant contribution of PSSs to designability of structures is   
  dominant in selection of HDSs. Even in  those regions where the
  designability is much larger than constant   
  part, there is still a strong correlation   
  between designability and its constant part. Thus those
  structures, which are attractive for PSSs, are in the set of
  HDSs. This suggests a close relation    
  between average energy gap of a structure and its constant part of  
  designability.    
  
  We find two geometrical necessary conditions for a structure to be HD. The   
  first one is compactness. This is because the PSSs select only highly 
  compact structures as absolute native states. Also we find that all  
  HDSs have a non-degenerate   
  NV representation. In average the designability of   
  structures decreases by increasing the degeneracy of neighborhood vector   
  and decreasing of compactness. This result shows that two monomer type pair  
  contact models have the common result for designability of HDSs with the  
  solvation model which is consistent with the recent study \cite{Buchler2}.  
  The relevant question remains   
  what will happen in models with more than two monomer types.
  
  {\bf Acknowledgment:}    
   We would like to thank N. Hamedani,
  R. Golestanian, N. Buchler, for helpful discussions. We also thank
  R. Everaers for carefully reading the manuscript and useful comments.

\end{multicols}  
  \newcommand{\PNAS}[1]{ Proc.\ Natl.\ Acad.\ Sci.\ USA\ {\bf #1}}   
  \newcommand{\JCP}[1]{ J.\ Chem.\ Phys.\ {\bf #1}}   
  \newcommand{\PRL}[1]{ Phys.\ Rev.\ Lett.\ {\bf #1}}   
  \newcommand{\PRE}[1]{ Phys.\ Rev.\ E\ {\bf #1}}   
  \newcommand{\JPA}[1]{ J.\ Phys.\ A\ {\bf #1}}

  \newpage  
  \begin{center}  
    \Large  
    Figure Captions \\[10mm]  
    \normalsize  
  \end{center}  
  
  {\bf Figure 1.}  
  
  \parbox{16cm}{The Average Designability of all native   
    structures against interaction parameters $E_c$  
    and $\gamma$ in a $10$ by $10$ square region with mesh $0.1$. The compact and   
    swollen regime correspond to areas of interaction space   
    with large and small average designability respectively.}  
  
  {\bf Figure 2.}  
  
  \parbox{16cm}{ Histogram of Designability for two pairs of interaction parameters.  
    a) $E_c=3$ and $\gamma=8$ (swollen regime), b)   
    $E_c=9$ and $\gamma=0.5$ (compact regime). }  
  
  {\bf Figure 3.}  
  
  \parbox{16cm}{The histogram of the number of structure with average ranks inside the  
    intervals with width 50 is compared with the uncorrelated (dash line)  
    and fixed (solid line) cases.}  
  
  {\bf Figure 4.}  
  
  \parbox{16cm}{ The Designability of structures {\it vs. } the number of PSSs which  
    choose the structure as unique ground state (the constant part of   
    designability) for two pairs of interaction parameters.  
    a) $E_c=3$ and $\gamma=8$ (swollen regime), b)   
    $E_c=9$ and $\gamma=0.5$ (compact regime).}


\begin{thebibliography}{40}   
  \bibitem{Anfinsen}{C.B. Anfinsen, Science {\bf 181}, 223 (1973).}   
  \bibitem{Chothia}{C. Chothia, Nature {\bf 357}, 543 (1992 ).}   
  \bibitem{Orengo}{C.A. Orengo, D.T. Jones, J.M. Thornton, Nature {\bf 372}, 631 (1994)}    
  \bibitem{Tang}{For a recent review, see C. Tang, Preprint   
      Cond-mat/9904377 to appear in Physica A}   
  \bibitem{Govindara}{S. Govindarajan, R. Recabarren, R.A. Goldstein, Proteins : Struct. Funct. Genetics {\bf 35}, 408 (1999)}   
  \bibitem{Hellinga}{H.W. Hellinga, \PNAS{94}, 10015 (1997).}  
  \bibitem{Li0}{H. Li, C. Tang, N.S. Wingreen, \PRL{79}, 765 (1997).}
  \bibitem{Li}{H. Li, R. Helling, C. Tang and N. Wingreen ,   
      Science {\bf 273}, 666 (1996).}   
  \bibitem{Chan}{H.S. Chan and K.A. Dill, \JCP{95}, 3775 (1991).}   
  \bibitem{Bornberg}{E. Bornberg-Baure, BioPhys. J. {\bf 73}, 2393 (1997).}   
  \bibitem{Goldstein}{S. Govindarajan, R.A. Goldstein, \PNAS{93}, 3341 (1996).}    
  \bibitem{Melin}{R. Melin, H. Li, N.S. Wingreen, C. Tang, \JCP{110}, 1252 (1999).} 
  \bibitem{Pande}{V.S. Pande, A.Yu. Grosberg and T. Tanaka,   
      \JCP{103}, 9482 (1995).}   
  \bibitem{Self1}{M.R. Ejtehadi, N. Hamedani, H. Seyed-Allaei,   
      V. Shahrezaei and M. Yahyanejad,   
      \PRE{57}, 3298 (1998); \JPA{31}, 6141 (1998).}   
  \bibitem{Kussel}{E.L. Kussell and E.I. Shakhnovich, \PRL{83}, 4437 (1999).}   
  \bibitem{Buchler}{N.E.G. Buchler, R.A. Goldstein, Proteins : Struct. Funct. Genetics {\bf 34}, 113 (1999).}   
  \bibitem{Li2}{H. Li, C. Tang, and N. Wingreen, \PNAS{95}, 4987 (1998).}   
  \bibitem{Buchler2}{N.E.G. Buchler, R.A. Goldstein, \JCP{112}, 2533 (2000).}   
  \bibitem{Self2}{M.R. Ejtehadi, N. Hamedani and V. Shahrezaei, \PRL{82}, 4723 (1999).}   
  \bibitem{Chan2}{H.S. Chan,   
      K.A. Dill, \JCP{90}, 492 (1989);   
      H.S. Chan, K.A. Dill, D. Shottle,   
      ``Statistical Mechanics and Protein Folding",   
      {\it  Princeton Lectures on Biophysics}, W. Bialek ed.,   
      (World Scientific, 1992).}   
  \bibitem{Self3}{V. Shahrezaei, N. Hamedani and M.R. Ejtehadi,
      \PRE{60}, 4629 (1999).}    
  \bibitem{Tatsumi}{R. Tatsumi and G. Chikenji, \PRE{60}, 4696 (1999).}     
  \bibitem{Shih}{C.T. Shih, Z.Y. Su, J.F. Gwan, B.L. Hao, C.H. Hsieh, and   
      H.C. Lee, \PRL{84}, 386 (2000).}        
    
  \end{thebibliography}
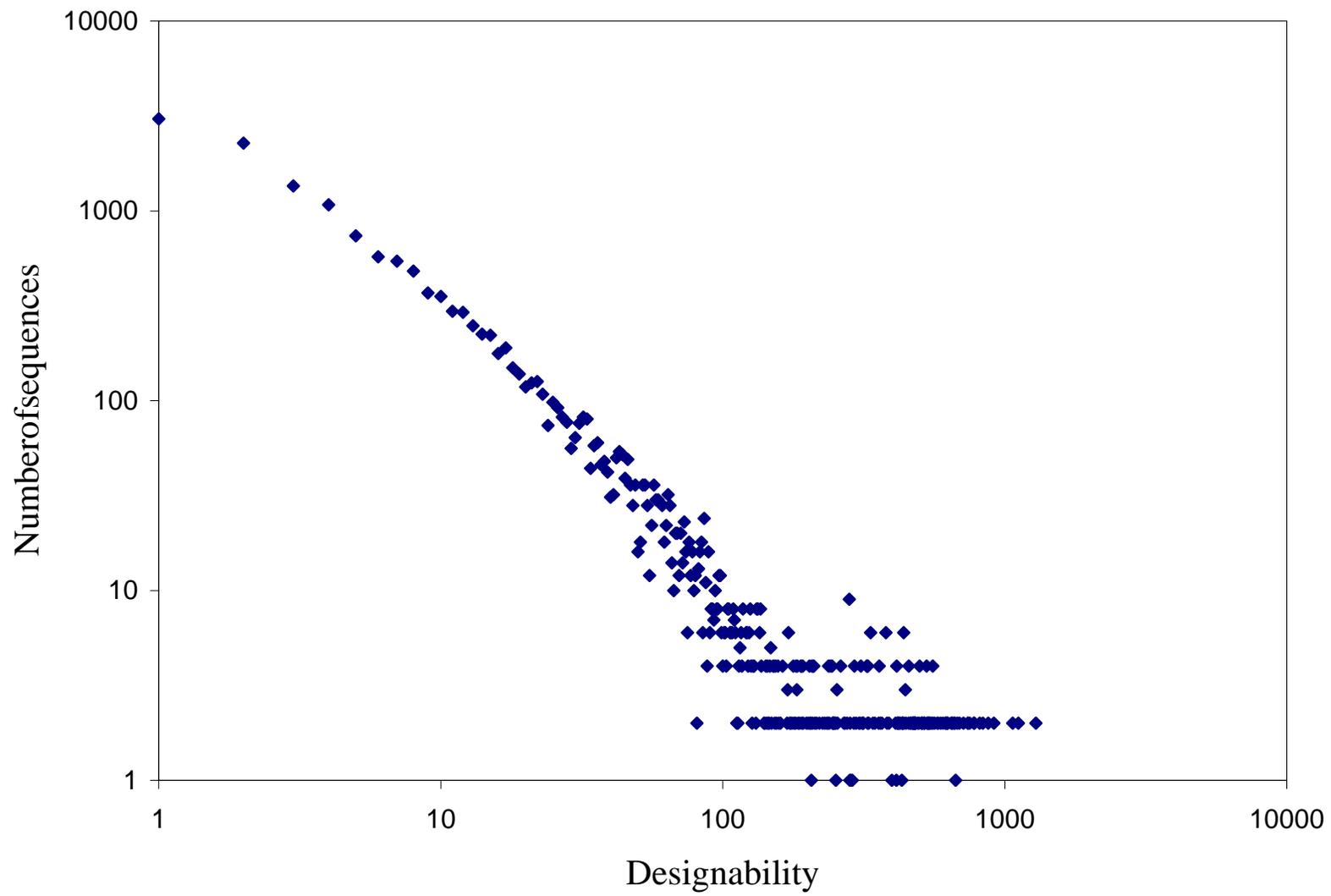
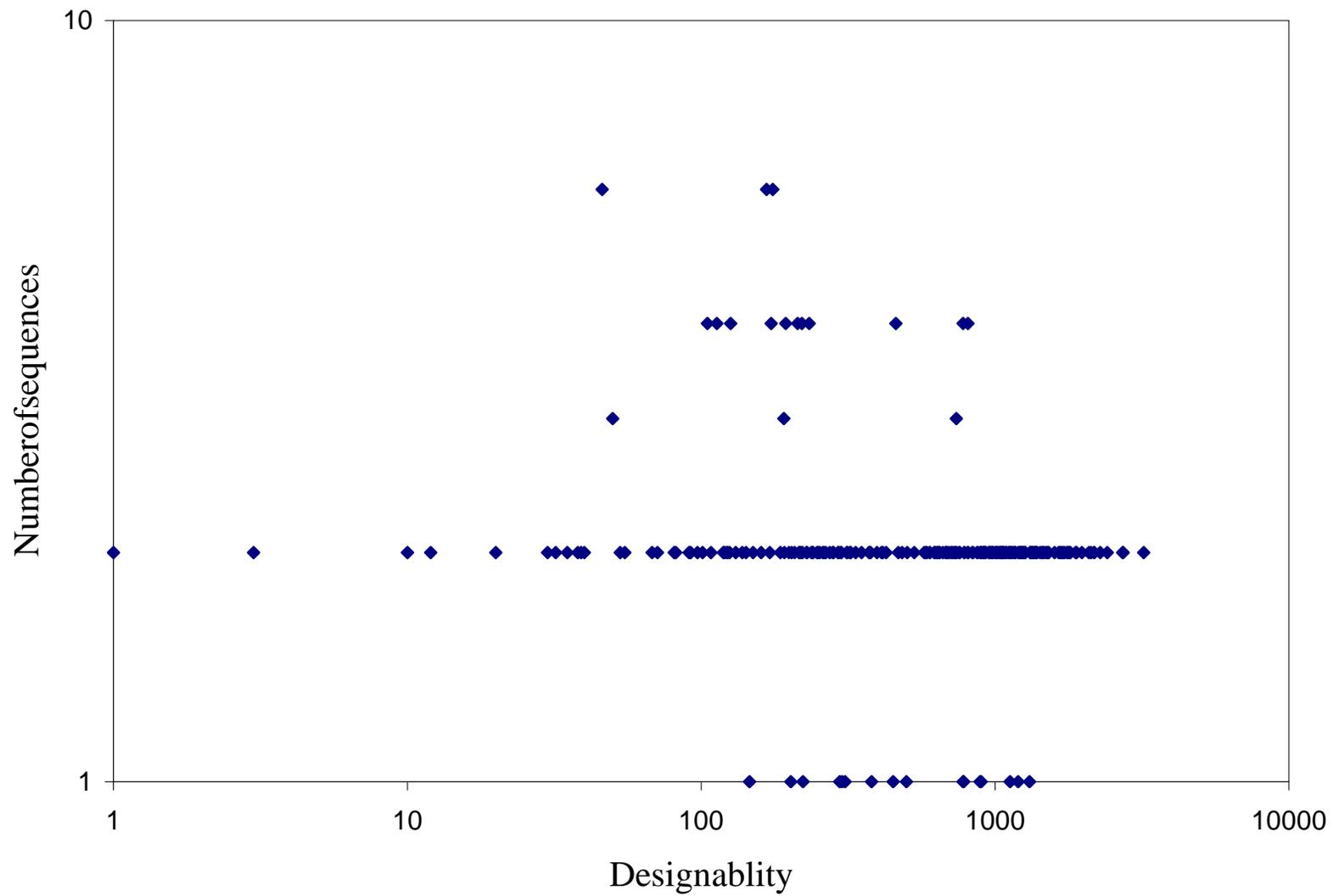
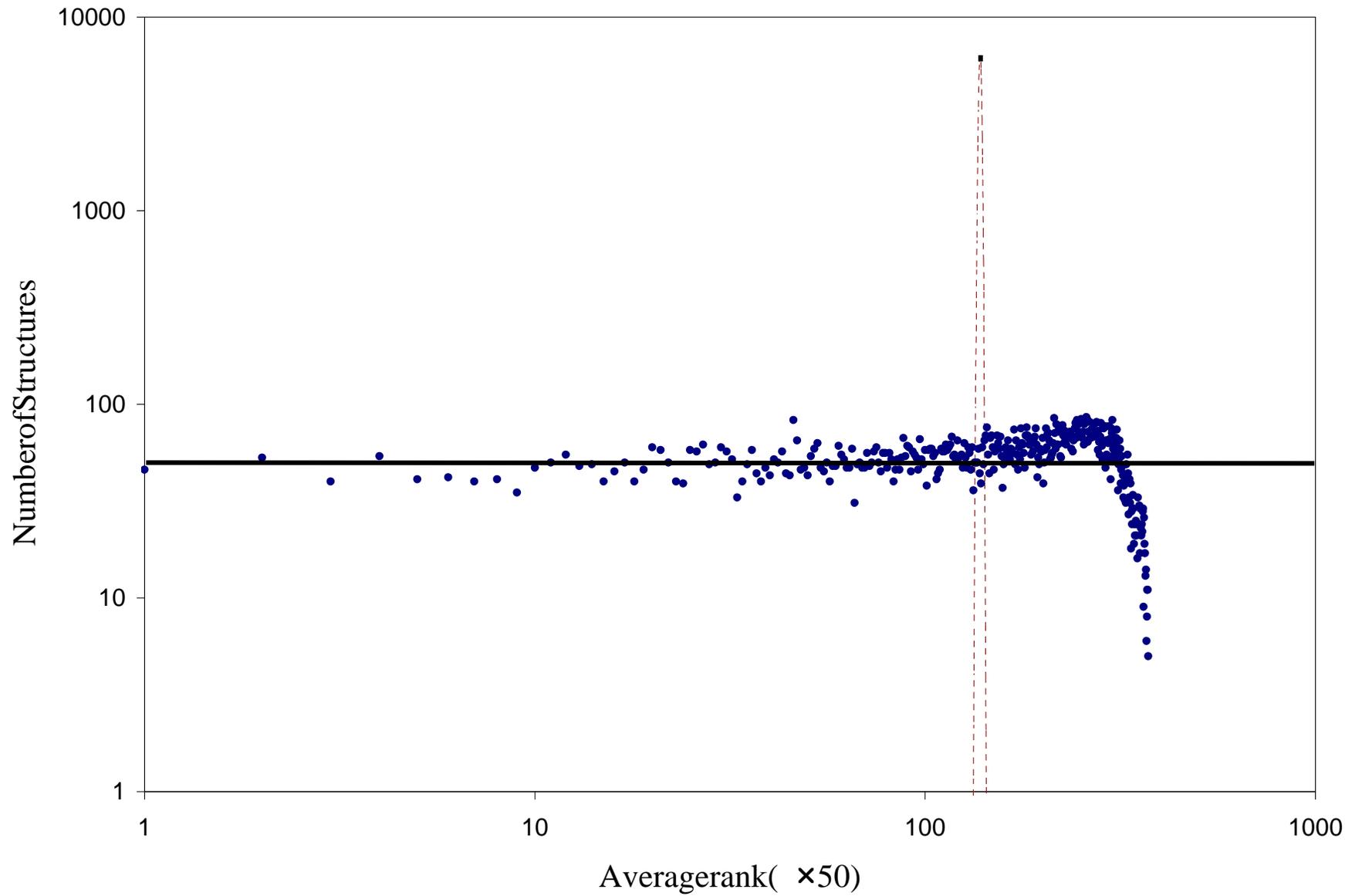
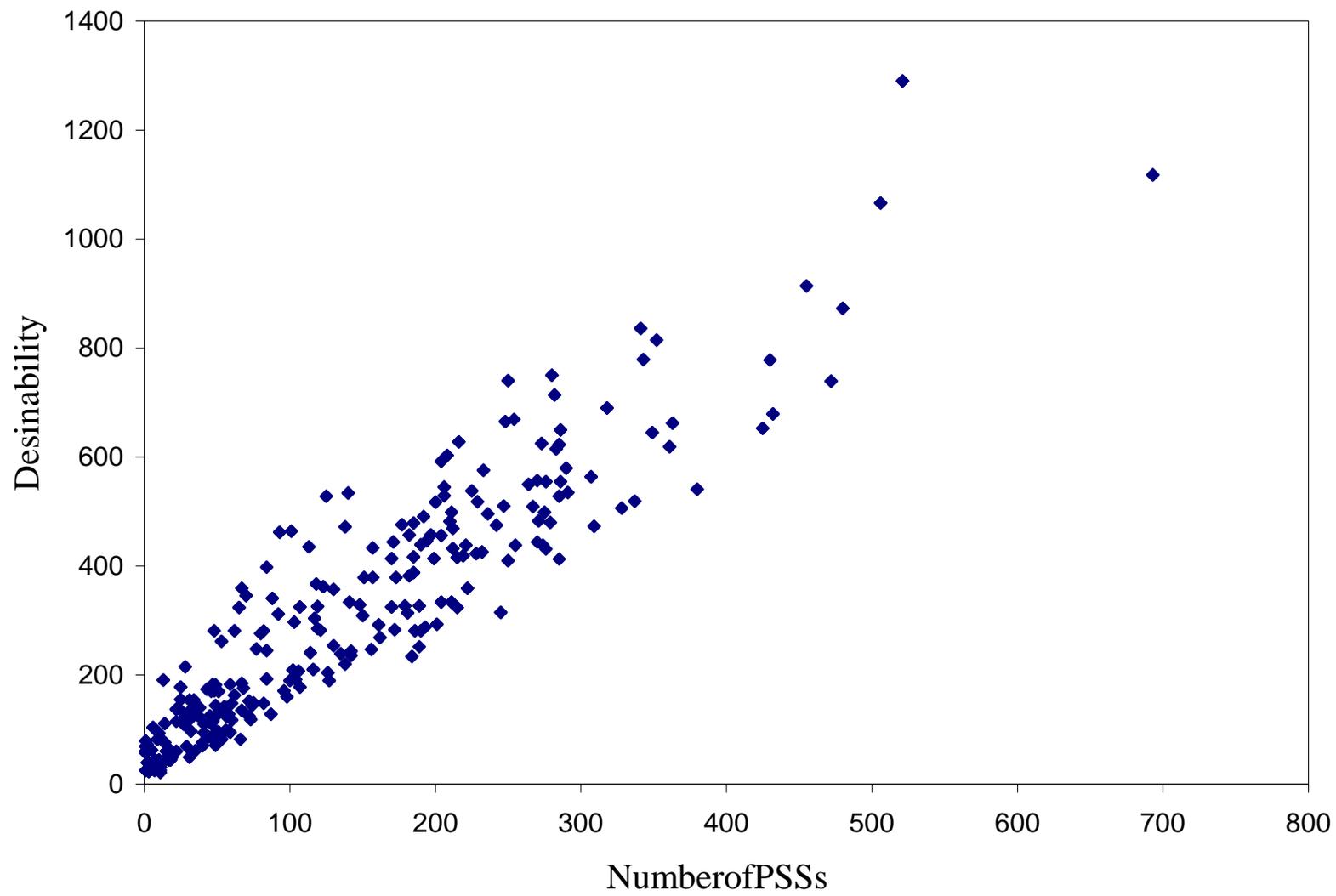
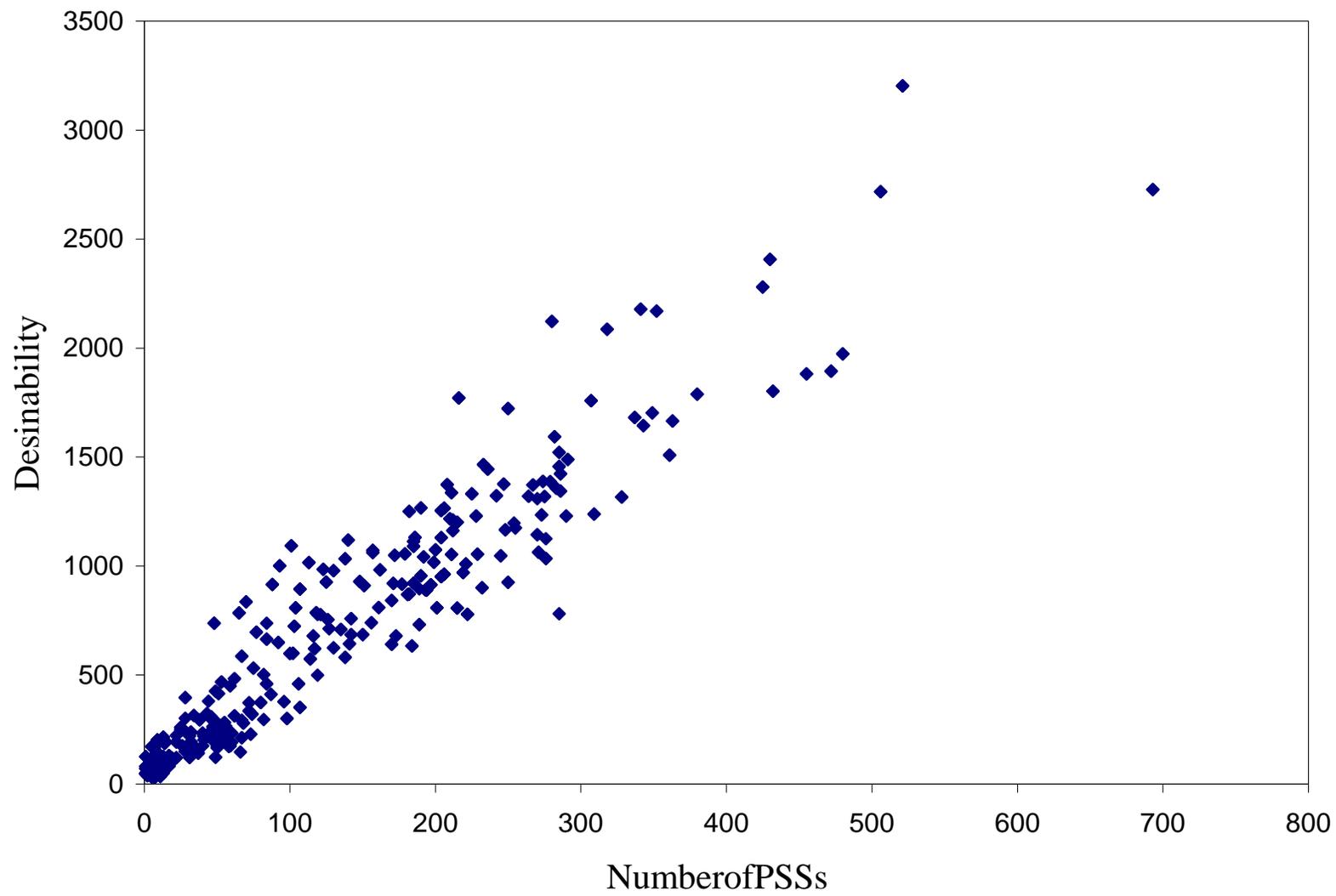
\end{document}